\documentclass[prd,superscriptaddress,amsfonts,amssymb,amsmath,showpacs,twocolumn,showkeys]{revtex4-1}

\usepackage{amsmath}
\usepackage{epsfig}
\usepackage{graphics}
\usepackage{color}
\usepackage{graphicx}
\usepackage[colorlinks=true,
      linkcolor=black,
      urlcolor=red,
      citecolor=blue]{hyperref}

\begin{document}


\title{Analytical model of dark energy stars}

\author{Ayan Banerjee}
 \email{ayan\_7575@yahoo.co.in}
\affiliation {Astrophysics and Cosmology Research Unit, University of KwaZulu Natal, Private Bag X54001, Durban 4000,
South Africa}
 
 \author{M. K. Jasim}
 \email{mahmoodkhalid@unizwa.edu.om}
\affiliation {Department of Mathematical and Physical Sciences, College of Arts and Science, University of Nizwa, Nizwa, Sultanate of Oman}

\author{Anirudh Pradhan}
 \email{pradhan.anirudh@gmail.com}
\affiliation {Department of Mathematics, Institute of Applied Sciences \& Humanities,
GLA University, Mathura-281 406, Uttar Pradesh, India}

\date{\today}

\begin{abstract}
In this article we study the structure and stability of compact astrophysical objects which are ruled by the dark energy equation of state (EoS). The existence of dark energy is important for explaining the current accelerated expansion of the universe.
Exact solutions to  Einstein field equations (EFE) have been found by considering particularized metric potential, Finch and Skea ansatz \cite{Finch}. The obtained solutions are relevant to the explanation of compact fluid sphere. Further, we have observed at the junction interface, the interior solution is matched with the Schwarzschild's exterior vacuum solution. Based on that, we have noticed the obtained solutions are well in agreement with the observed maximum mass bound of $\approx$ 2$M_{\bigodot}$, namely, PSR J1416-2230, Vela X-1, 4U 1608-52, Her X-1 and PSR J1903+327, whose predictable masses and radii are not compatible with the standard neutron star models.  Also, the stability of the stellar configuration has been discussed briefly, by considering the energy conditions, surface redshift, compactness, mass-radius relation in terms of the state parameter ($\omega$). Finally, we demonstrate that the features so obtained are physically acceptable and consistent with the observed/reported data \citep{Gondek-Rosinska,Glendenning}. Thus, the present dark energy equation of state appears talented regarding the presence of several exotic astrophysical matters.

\end{abstract}

\maketitle

\section{Introduction}
The present accelerated expansion of the universe has been established by  various self-governing high-precision observational data such as the galaxy rotation curves, Supernovae type Ia \citep{Grant,Perlmutter} and cosmic microwave background radiation \citep{Bennett,Hinshaw}. Several models have been proposed to account for this observed late-time  accelerated expansion of the Universe. One of the assumptions of the current paradigm of cosmology is the existence of so-called dark energy, where general relativity is assumed to be correct. The dark energy 
component is characterized by a negative pressure. A large volume of phenomenological and theoretical  models have been proposed to describe the actual situation (see \cite{Copeland} and references therein). In this regard, the simplest and most efficient way to describe the present situation  is by introducing a cosmological constant $\Lambda$ within general relativity (GR). Overall, DE is thought  to contribute 70\% of the worldwide energy budget in the universe. See Ref \cite{Li:2011sd} for a recent review on DE models. However, despite its success, $\Lambda$CDM  model, which consists of a cosmological constant ($\Lambda$) plus Cold Dark Matter (CDM), suffers the theoretical problems for instance  the \textit{fine-tuning} and \textit{cosmic happenstance}  mysteries. So, it is important to investigate other available theoretical schemes such as employing an EoS p = $\omega \rho$, where $p$ is the pressure and $\rho$ is the energy density, respectively. Without entering further details, the requirement for cosmic expansion is  $\omega <-1/3$ , as shown through Friedmann equation $\ddot{a}/a$ = -$4 \pi G(p + \rho/3)$, whereas $\omega=-1$  reduces to the particular case of cosmological constant. On the other hand, the parameter range $-1< \omega<-1/3$ is called quintessence models and the dark energy is reducing accordingly  with a scale feature $a(t)$ as $\rho_Q \equiv a^{-3(1+\omega)}$ \cite{Turner}. In \cite{Alam:2003sc}, authors have shown that depending on the evolution of 
density perturbations one can distinguish a cosmological constant ($\omega=-1$) from quintessence models with $\omega \geqslant - 0.9$  at the 3$\sigma$ level. It is therefore the equation of state parameter violates the null energy condition, i.e. $\omega_{de}< -1$. However, the EoS of a phantom scalar field is always less than $-1$.  

Our interest is to search for local astrophysical objects within this scenario.  In this line of thought, gravastar (gravitational vacuum star) model received special attention - as an alternative to a black hole solution has been proposed by Mazur and Mottola \cite{Mazur}. 
This model consist of five layers including two thin--shells. The de-Sitter geometry in the interior governed by an EoS p = -$\rho$, matches to an exterior Schwarzschild solution.
It is argued that in between the interior and exterior geometry, there is a finite thickness shell with an equation of state p = +$\rho$. The shell is comprised of stiff fluid matter. See Ref \cite{Visser} for a recent development on gravastar model where five-layer model can be simplified to three-layer with the phase transition layer was replaced by a single spherical $\delta$-shell. In Ref \cite{Bilic:2005sn}
an argument claiming for alternative model of gravastar where the de-Sitter regime was replaced by an interior solution governed by a Chaplygin gas equation of state, interpreted as a Born-Infeld phantom gravastar. We refer to our reader 
for a recent discussion see e.g., \cite{Wiltshire}. 

Apart from the aforementioned gravastar solution, one may consider extended objects supported by
dark energy EoS (interior region). The motivation comes from the fact that dark energy exerts a repulsive force on its surrounding,  and this repulsive force may prevent the star from collapsing. This model found great success and denoted as a \textit{dark energy star} \cite{GChapline}. After this work, Lobo \cite{lobo1}
generalized of the gravastar picture by considering the dark energy EoS, $\omega= p/ \rho <-1/3$,  which is matched to the exterior Schwarzschild geometry at a junction interface. According to the proposed model large stability regions do exist, but sufficiently close to where the event horizon is expected to form.
The results of this analysis had led to another possibility that massive star doesn't simply collapse to form a black hole.  Furthermore, Chaplygin dark star was proposed in \cite{Bertolami:2005pz}.  A related simplified model with time-dependent spacetime was proposed in Ref. \cite{DeBenedictis:2008qm} with an evolving parameter $\omega$ crossing the phantom divide, $\omega$ = -1. Various kinds of dark stars were found to be stable under small perturbations (See in detail in Refs. \cite{Lobo:2006ue}).

In \cite{Yazadjiev:2011sm}, Yazadjiev found a class of exact interior solutions describing mixed relativistic stars. According to the model dark energy was provided by scalar fields with negative kinetic energy. Whereas the dark energy imprints in gravitational wave spectrum of mixed neutron-dark-energy stars (containing both ordinary matter and dark energy) have been found in the literature, and we refer the reader to Ref \cite{Yazadjiev:2011sd}. Lately, an interest in G-lump - a vacuum self gravitation particle-like structure without horizons was introduced by Dymnikova \cite{Dymnikova:2001fb}.

Motivated by the undergoing a phase of accelerated expansion of our universe which was confirmed by extremely  luminous stellar explosions, known as type Ia supernovae (SNeIa), provokes us to rethink the commonly accepted scenario. Another related issue is to assess the properties of neutron stars and there formation in our modern scenarios, which is changed after the discovery of PSR J1614-2230 \cite{Demorest} as 1.97 $\pm$ 0.04 $M_{\odot}$. This discovery puts a severe constraint on the EoS for nuclear matter. In particular, one needs  to consider the appearance of exotic particle at densities $\sim$ 5 - 8 $\times 10^{14}$ g/cm$^3$ \cite{Nagae}. Nevertheless, ultra-luminous X-ray sources (ULXs) are X-ray sources with  luminosity above the Eddington limit cannot be explained by the conventional idea of normal stellar mass black hole. Therefore,  the basic nature of ultraluminous X-ray pulsars are \citep{pcc,mlr}, as of now,  remains unsolved \cite{Bachetti:2015pwa}. However, different models have been suggested for the ULX-pulsars including magnetic field of different strength, but none of them are conventional one. With this viewpoint, 
we explore dark energy as a possible source which could constraints on the mass-radius relation for compact objects whose estimated masses and radii are not compatible with our known sources,  such as X-ray burster 4U 1820-30, X-ray sources PSR J1614 - 2230 and X-ray pulsar Vela X - 1, Cen X - 3.

On the other hand, the effect of anisotropy on the evolution of self gravitating objects has been studied extensively by several authors using both analytical and numerical methods.
The anisotropy means that there are two components of pressure, one is radial direction and the other is along the transverse directions (see Ref. \cite{Bowers,Ruderman} and the references within).  Interestingly, the interior solution for anisotropy fluid distribution can arise from various reasons: the presence of type 3A superfluid phase, a mixture of different types of fluids or the presence of a magnetic field, etc. (see Ref. \cite{Kippenhahm,Sokolov}). From recent investigations of the stellar structure,  Herrera and Barreto \cite{Herrera} considered polytropes for anisotropic matter both in the Newtonian and the GR regimes \cite{Herrera1,Herrera2}. In \cite{Feroze,Maharaj2}, charged anisotropic solutions with a quadratic equation of state was obtained. In particular static anisotropic fluid spheres with uniform energy density was considered in \cite{Maharaj1}. Recently, Mak and Harko \cite{Mak}, showed that anisotropy  may play a vital role in the stability of a dense stars with strange matter. Other studies are also available (check references for instance  \cite{Maurya:2018kxg}). Through this suggestions one could look for anisotropic dark energy model at small scales, and hence hope for yet another approach to find other possible solution satisfying the physical criteria and necessarily contains.

This paper is organized as follows. In Sec.~\ref{sec2}, the EFEs have been developed assuming a specific choice of interior space-time (Finch and Skea type), which is static and spherically symmetric. Then, the dark energy EoS have been used to present the structural equations of dark energy star and verify our solutions with compact star Vela X-1. In Sec.~\ref{sec3}, the matching of interior space-time to the exterior Schwarzschild vacuum solution at a junction interface have been studied. In Sec.~\ref{sec4}, we have explored physical properties of compact object for known mass - radius and then determine the mass-radius relation and compactness of star for different compact objects. Finally, a brief discussion is given in Sec.~\ref{sec5}.

\section{Basic concepts of Einstein's field equations (EFE)}\label{sec2}
The interior metric of a static and spherically symmetric star solution is defined as 
\begin{eqnarray}\label{eq1}
ds^{2}&=& -exp\left[{-2\int_r^{\infty}g(\tilde{r})d\tilde{r}}\right]dt^{2} +\frac{dr^{2}} {1-\frac{2m}{r}} \nonumber \\
&\;& +r^{2}(d\theta^{2}+\sin^{2}\theta d\phi^{2}),
\end{eqnarray}
where the functions $g(r)$ and $m(r)$ are random functions of the radial coordinate $r$. The factor $g(r)$ is locally measured for gravitational acceleration. As a consequence, positive and negative $g(r)$ means inwardly gravitational attraction, and an outward gravitational repulsion, respectively. 

The matter contained in the spherical object is described by the anisotropic fluid, and the corresponding energy-momentum tensor will be then 
\begin{equation}\label{eq2}
T_{\mu\nu}=(\rho+p_t)U_{\mu}U_{\mu}+p_t g_{\mu\nu}+(p_r-p_t)\chi_{\mu}\chi_{\nu},
\end{equation}
where $U_{\mu}$ represents the 4-velocity of the fluid and $\chi_{\mu}$ is the unit 4-vector
along the radial direction. As a matter source in (\ref{eq2}), we can also justify by summoning the diagonal components of stress-energy tensor
$T^{\mu} _\nu$ = diag~[−$\rho$, $p_r$, $p_t$, $p_t$]. The GR field equations for 
the metric (\ref{eq1}) with the following energy-momentum tensor (\ref{eq2}) can be written as
\begin{eqnarray}\label{eq3}
m'(r) &=& 4\pi r^{2}\rho,  \\ \label{eq4}
g(r) &=& \frac{m+4\pi r^{3}p_r}{r(r-2m)},  \\
p_r'&=& -\frac{(\rho+p_r)(m+4\pi r^{3}p_r)}{r(r-2m)}+\frac{2}{r}(p_t-p_r) \label{eq5},
\end{eqnarray}
where `prime' denotes the differentiation with respect to $r$. Along these lines, $\rho(r)$ is the energy density, whereas $p_r(r)$ and $p_t(r)$ are
the radial and transverse pressures, respectively. Eq. (\ref{eq5}), corresponds to the Bianchi's identity suggesting that (the covariant conservation of the stress-energy tensor) $\Delta_\mu T^{\mu\nu}=0$. By considering relativistic Tolman-Oppenheimer-Volkoff (TOV) equation, one can also achieve the conservation equation.

Since, the system of Eqs. (\ref{eq3}-\ref{eq5}) having three equations with five unknowns,
namely, $\rho$(r), $p_r$(r), $p_t$(r), $g(r)$ and $m(r)$. Thus, the unknown functions have to solve for achieving all the plausible features of a compact star. To find the solution of these differential equations, we assume a metric potential \textit{ansatz}, namely, Finch and Skea type \cite{Finch}. The metric potential function $e^{\lambda}$ is given by the equality:
\begin{equation}\label{eq6}
e^{\lambda}=\left(1-\frac{2m}{r}\right)^{-1} =1+\frac{r^{2}}{R^{2}},
\end{equation}
where $R$ is related with curvature parameter of the configuration which satisfy all physical criteria for a stellar source.  Now, solving (\ref{eq6}) one can derive the mass function
\begin{equation}\label{eq7}
m(r)=\frac{r^{3}}{2(R^{2}+r^{2})}.
\end{equation}
In next, we adopt the dark energy EoS, $p_r$ = $\omega \rho$ with $\omega < 0$, which is one of the more promising direction to elucidate the current accelerated expansion of the universe. Recently, Ghezzi \cite{Ghezzi} found a compact object coupled to inhomogeneous anisotropic dark energy.  Moreover, in essence of Ref. \cite{Bertolami}, Bertolami and Paramos studied spherically symmetric dark energy structure using a polytropic EoS of negative index.
Dark stars were further extended that describe the collapse of a spherical object from an initial
state of positive pressure to a final state with negative pressure inside a finite radius core \cite{Beltracchi:2018ait}.

\begin{figure}[h!]
\begin{center}
\includegraphics[width=8.cm]{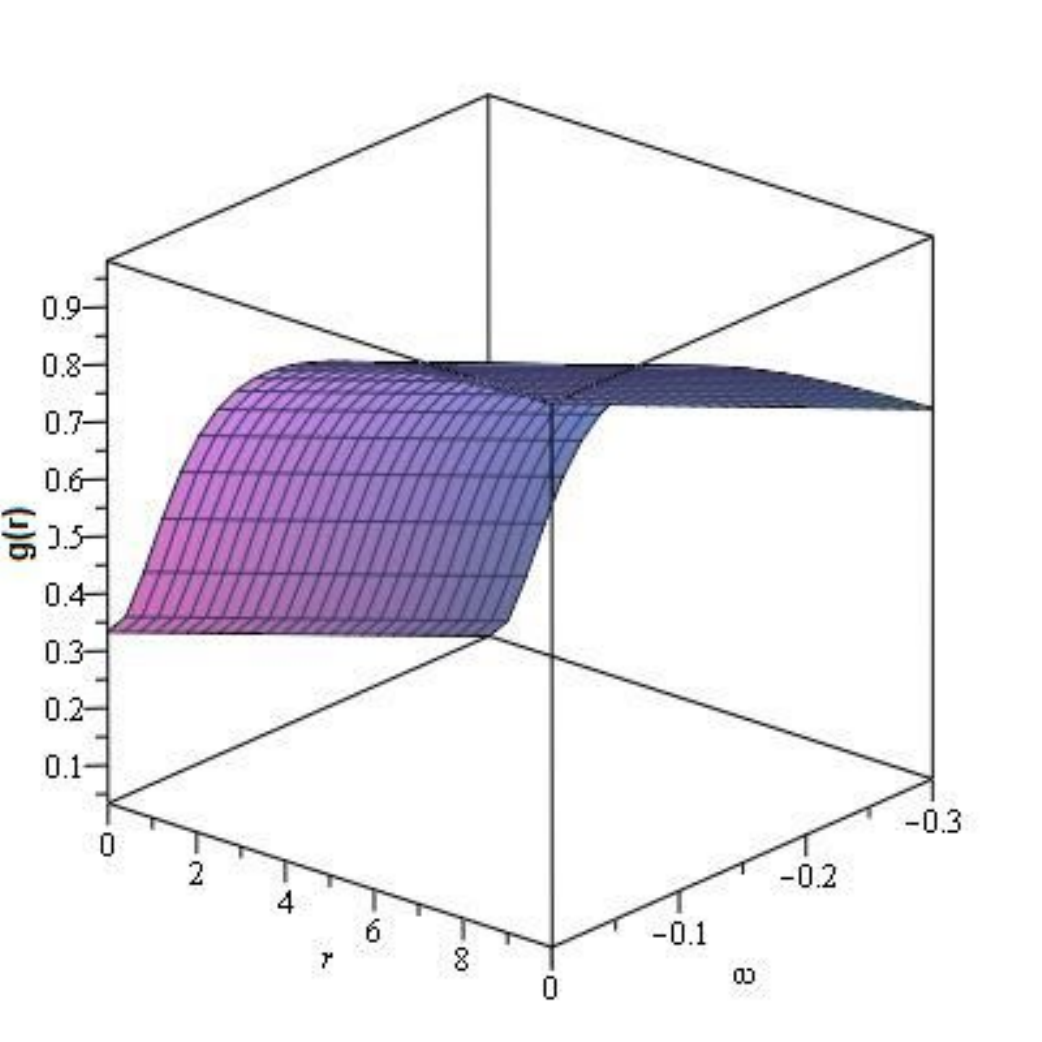} \includegraphics[width=8cm]{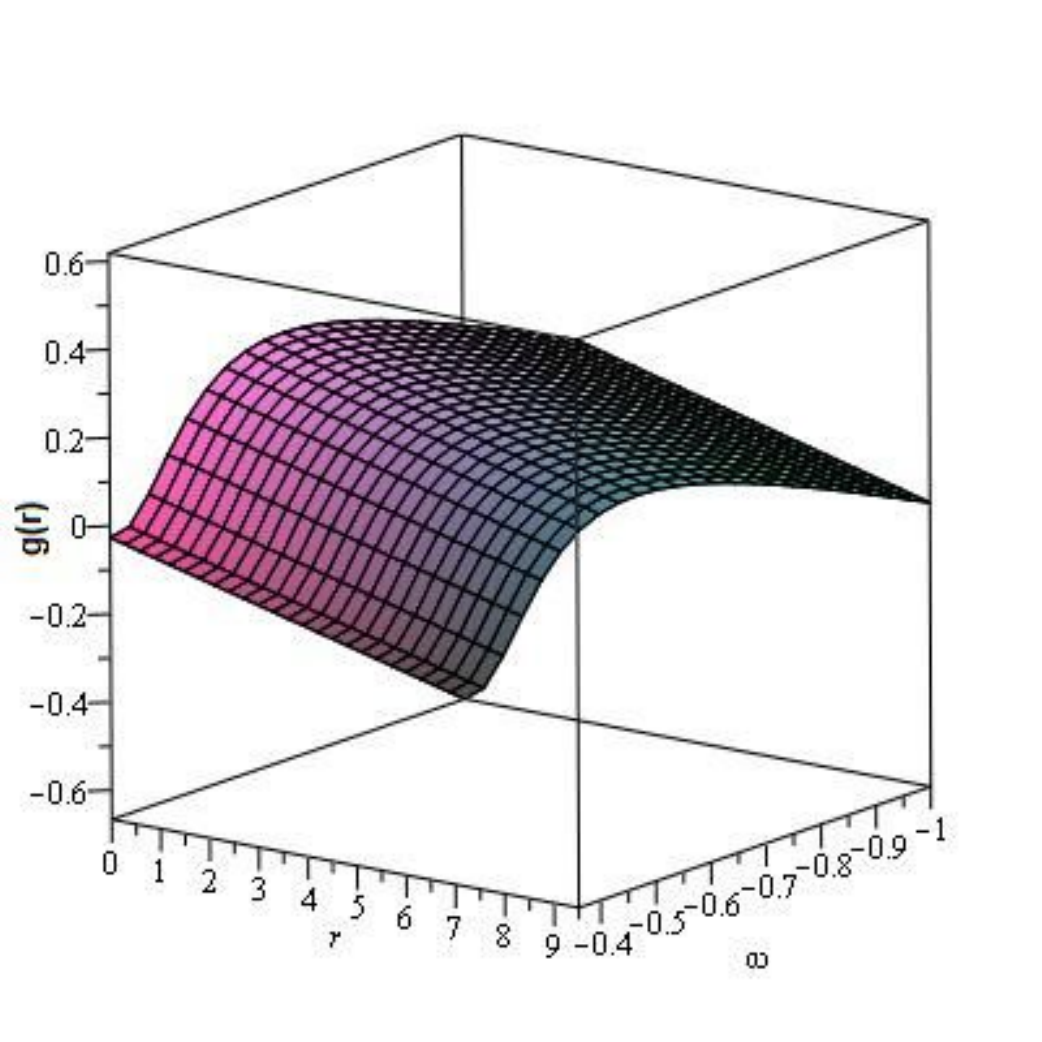}
\caption{The ``gravity profile", $g(r)$, is plotted within the interval for $-1/3<\omega<0 $, and then for $-1<\omega<-1/3 $ respectively on the top and bottom panels. Analysis shows that $g(r)$ takes positive value i.e. indicating an inwardly gravitational attraction for $-1/3<\omega<0 $. To be a solution of a dark star we necessarily exclude this region. }\label{f7}
\end{center} 
\end{figure}

  Now, taking into account the Eqs. (\ref{eq3}) and (\ref{eq4}), and using the above EoS we obtain the function $g(r)$ as follows:
\begin{equation}\label{eq8}
g(r)=\frac{r}{2R^{2}}\left\{1+\frac{\omega(3R^{2}+r^{2})}{R^{2}+r^{2}}\right\},
\end{equation}
then the Eqs. (\ref{eq3})-(\ref{eq5}) reduce to a simple system of algebraic equations
\begin{eqnarray}
\rho &=& \frac{1}{8\pi}\frac{3R^{2}+r^{2}}{(R^{2}+r^{2})^{2}},\label{eq9}\\
p_r &=& \frac{\omega}{8\pi}\frac{3R^{2}+r^{2}}{(R^{2}+r^{2})^{2}}, \label{eq10}\\
p_t &=&\frac{1+\omega}{32\pi}\frac{3R^{2}+r^{2}}{(R^{2}+r^{2})^{2}}\frac{r^{2}}{R^{2}}
\left\{1+\omega\frac{3R^{2}+r^{2}}{R^{2}+r^{2}}\right\}  \nonumber \\
&\;& -\frac{\omega}{8\pi}\frac{R^{2}(r^{2}-3R^{2})}{(R^{2}+r^{2})^3}, \label{eq11}
\end{eqnarray}
where $\Delta$ = $p_t - p_r$  stands for measure of the pressure anisotropy of the fluid comprising the dark energy star, can be expressed in the following form
\begin{equation}\label{eq12}
\Delta=\frac{1+\omega}{32\pi}\frac{3R^{2}+r^{2}}{(R^{2}+r^{2})^{2}}\frac{r^{2}}{R^{2}}
\left\{1+\omega\frac{3R^{2}+r^{2}}{R^{2}+r^{2}}\right\}-\frac{\omega r}{4\pi}\frac{5R^{2}+r^{2}}{(R^{2}+r^{2})^{3}}.
\end{equation}
It is known that $\Delta$ represents a force due to the anisotropic nature of the stellar model. The force is being directed outward when $\Delta$ $>$ 0 i.e. $p_t> p_r$ and inward when $\Delta$ $<$ 0 i.e. $p_t < p_r$. Corresponding to $\Delta$ = 0 is a  certain case of an isotropic pressure. At the stellar centre, $r = 0$ we have
$\Delta$ = 0, which is expected. Indeed, the central density is a non-zero constant, $\rho_0$= 3/8$\pi R^2$.
This reflects that there is no singularity inside the star. Our intention is to study and analyze the dark energy star that can be considered as a suitable way to explain those massive stellar systems like white dwarfs, massive pulsars and magnetars etc.

The main assumption that leads to Eq. (\ref{eq8}) is that $g(r) >$ 0, for $\omega > -(r^2+R^2)/(r^2+ 3R^2)$ indicating an inwardly gravitational attraction. The condition is verified from Fig. \textbf{1}, that $g(r)$ is positive in the interval $-1/3<\omega< 0$, while $g(r) < 0$ in the interval $ -1<\omega<-1/3$. The above discussions show that to be a gravastar like solution, the local acceleration due to gravity of the interior solution be repulsive.  In spite of these constraints, we consider only the region where $g(r) < 0$ for further precision (see Refs. \cite{lobo1} for more details).

Specifically, we investigate the maximum mass of a dark star using the dark energy equations of state $\omega <-1/3$, and study the effect of the state parameter $\omega $ on the other physical properties such as the density, mass-radius and gravitational redshift. In addition, we compare the obtained results of this theory with some strange/compact star candidates like PSR J1416-2230, Vela X-1, 4U 1608-52, Her X-1 and PSR J1903+327, respectively. However, the X-ray pulsar Vela X-1, whose estimated mass and radius are 1.77$\pm$ 0.08 $M_{\bigodot}$ and $R = 10.852$ Km has been allowed for testing the physical acceptability of the developed model.

\begin{figure}[h!]
\begin{center}
\includegraphics[width=8cm]{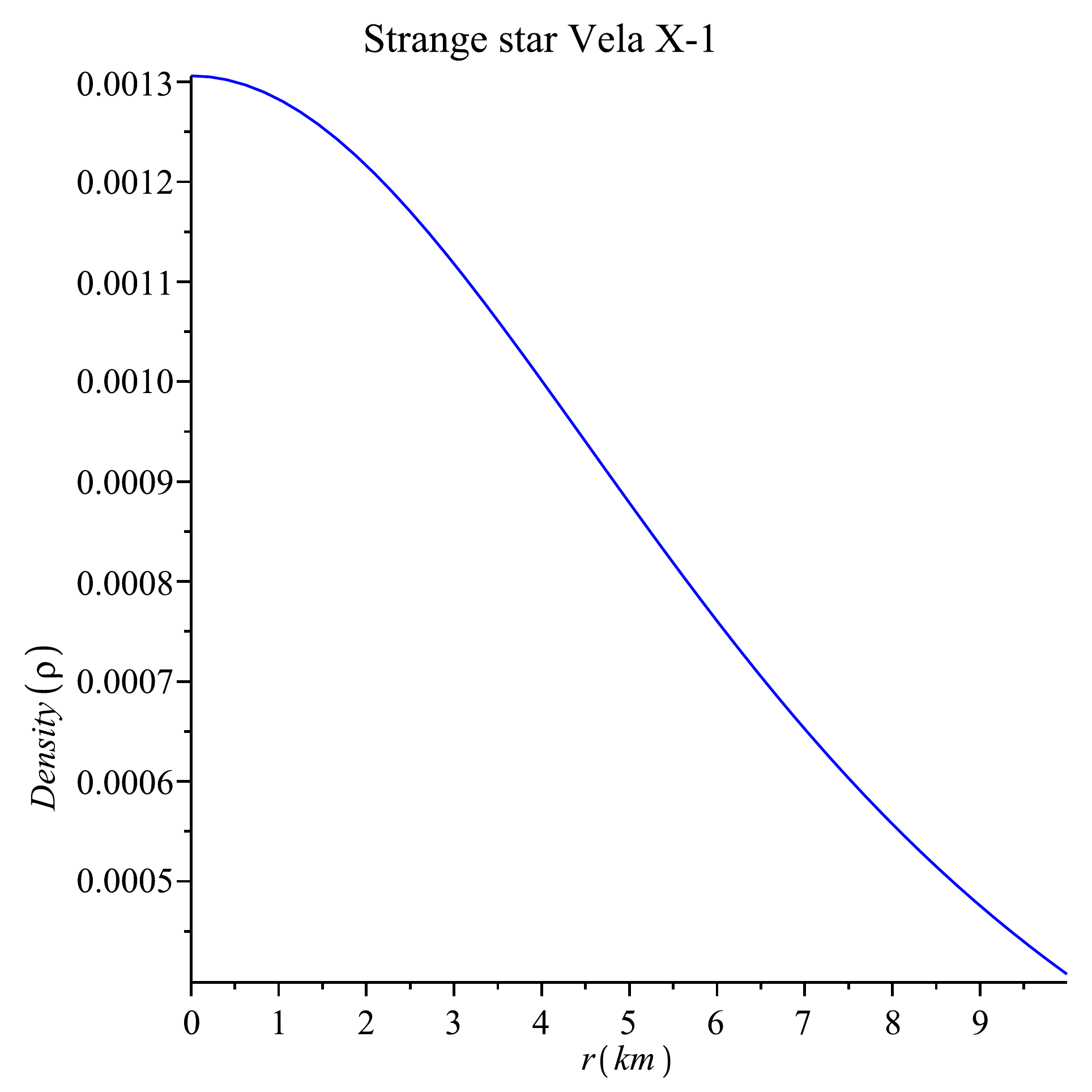} \includegraphics[width=8cm]{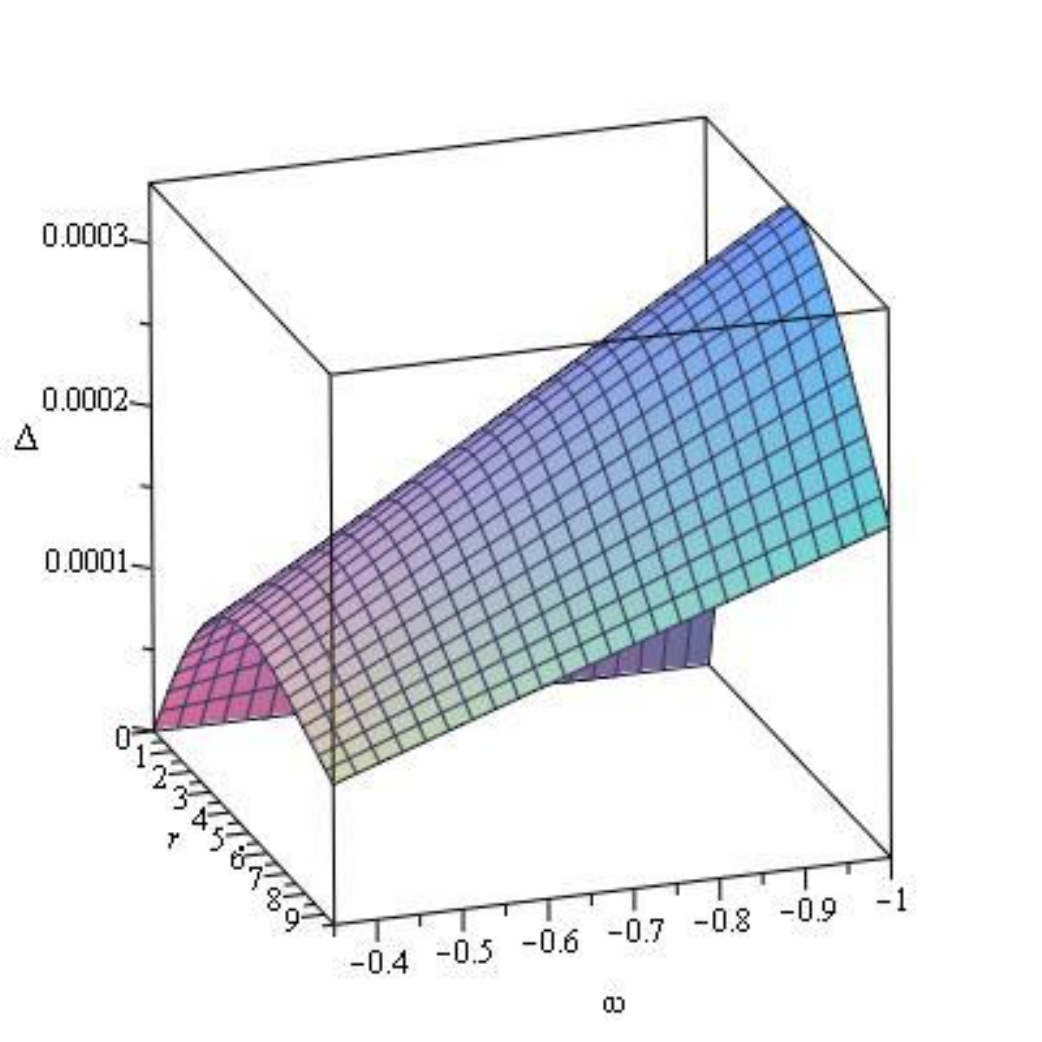}
\caption{Variation of the density and anisotropy $\Delta$ have been plotted as a function of the radial coordinate $r$.
For density profile we consider $\omega = - 0.35$ and radii are on the order of 10 km. For anisotropy factor,  we verify that $\Delta >0$ in the range $-1<\omega<-1/3 $.}\label{f7}
\end{center} 
\end{figure}

\section{Junction Condition }\label{sec3}
At this stage, the interior solution is matched  to an exterior Schwarzschild vacuum solution with $p$ = $\rho$ = 0 at a junction interface $\Sigma$, with junction radius, $a$.
The Schwarzschild exterior solution is given by
\begin{equation}\label{eq13}
ds^{2}=-\left(1-\frac{2M}{r}\right)dt^{2}+\frac{dr^{2}}
{1-\frac{2M}{r}}+r^{2}(d\theta^{2}+\sin^{2}\theta d\phi^{2}),
\end{equation}
which possesses an event horizon at $r_h = 2M$. We have chosen the value of $a > r_h$, to avoid the
presence of horizons i.e. the junction radius lies outside $2M$. 

Let us turn now our attention in computing the surface stress-energy tensor $S_{ij}$, which can be expressed in terms of the jump of the extrinsic curvature by means of the Lanczos equation \cite{Israel} (see Ref. \cite{Lobo:2005zu}  for more details) and defined by
\begin{equation}\label{eq14}
S^{i}_{j} = -\frac{1}{8\pi}\left(\kappa^i_j-\delta^i_j\kappa^m_m\right),
\end{equation}
where $k_{ij}$ represents discontinuity in the extrinsic curvature $K_{ij}$, across the junction interface, which is defined by
$k_{ij} = K^{+}_{ij}-K^{-}_{ij}$. Therefore, the second fundamental form has been used to yields the extrinsic curvature in its final form, which is
\begin{equation}\label{eq15}
K^{\pm}_{ij} = -\eta_{\nu}\left(\frac{\partial^2 x^{\nu}}{\partial\xi^i \partial\xi^j}+\Gamma^{\nu\pm}_{\alpha\beta}\frac{\partial x^{\alpha}}{\partial\xi^{i}}\frac{\partial x^{\beta}}{\partial\xi^{j}}\right),
\end{equation}
where $\eta_{\nu}$ represents the unit normal at the junction, the symbol $`\pm'$ represents the interior and exterior spacetime, and $\xi^i$ represents the intrinsic coordinates. Now, by using the metrics (\ref{eq1}) and (\ref{eq13}), the non-trivial
components of the extrinsic curvature are given by
\begin{equation}\label{eq16}
K_{\tau}^{\tau~+}=\frac{\frac{M}{a^{2}}+\ddot{a}}{\sqrt{1-\frac{2M}{a}+\dot{a}^{2}}},
\end{equation}
\begin{equation}\label{eq17}
K_{\tau}^{\tau~-}=\frac{\frac{a}{2}\left\{\frac{1}{R^{2}+a^{2}}+\omega\frac{3R^{2}+a^{2}}{(R^{2}+a^{2})^{2}}
\right\}+\ddot{a}-\frac{(1+\omega)a}{2R^{2}}\frac{3R^{2}+a^{2}}{R^{2}+a^{2}}\dot{a}^{2}}{\sqrt{\frac{R^{2}}{R^{2}+a^{2}}+\dot{a}^{2}}},
\end{equation}
and
\begin{eqnarray}\label{eq18}
K_{\theta}^{\theta~+}&=& \frac{1}{a}\sqrt{1-\frac{2M}{a}+\dot{a}^{2}}, \\ \label{eq19}
K_{\theta}^{\theta~-} &=& \frac{1}{a}\sqrt{\frac{R^{2}}{R^{2}+a^{2}}+\dot{a}^{2}}.
\end{eqnarray}
We have adopted the usual notation in which 
the dot and prime represent $d/d\tau$ and $d/dr$, respectively. Since our metrics are diagonal,  $S^{i}_{j}$ is also diagonalized and written as $S^{i}_{j}= \text{diag} ~(-\sigma,\mathcal{P},\mathcal{P})$. Thus, the Lanczos equations give the energy density and pressure on the shell:
\begin{equation}\label{eq20}
\sigma=-\frac{1}{4\pi a}\left[\sqrt{1-\frac{2M}{a}+\dot{a}^{2}}-\sqrt{\frac{R^{2}}{R^{2}+a^{2}}+\dot{a}^{2}}\right],
\end{equation}
\begin{eqnarray}\label{eq21}
\mathcal{P}&=&\frac{1}{8\pi a}\left[\frac{1-\frac{M}{a}+\dot{a}^{2}+a\ddot{a}}{\sqrt{1-\frac{2M}{a}+\dot{a}^{2}}}
-\frac{1+\frac{\omega a^{2}}{2}\frac{3R^{2}+a^{2}}{(R^{2}+a^{2})^{2}}}{\sqrt{\frac{R^{2}}{R^{2}+a^{2}}+\dot{a}^{2}}}
\right.\nonumber\\
&& \left.
-\frac{\dot{a}^{2}+a\ddot{a}-\frac{a^{2}}{2(R^{2}+a^{2})}
+(1+\omega)\frac{a^{2}}{2R^{2}}\frac{3R^{2}+a^{2}}{R^{2}+a^{2}}\dot{a}^{2}}
{\sqrt{\frac{R^{2}}{R^{2}+a^{2}}+\dot{a}^{2}}}\right],
\end{eqnarray}
where $\sigma$ and $\mathcal{P}$ are the surface energy density and surface pressure at the junction interference.

Here, we interpret the quantities $\sigma$ and $\mathcal{P}$ arising from the thin-shell formalism for the spherically symmetric case. In this regard, the equation of motion for the surface stress-energy tensor is given by $S^{i}_{j|i}$= $\left[T_{\mu\nu}e^{\mu}_{(j)}n^{\nu}\right]^{+}_{-}$, where $[X]^{+}_{-}$ denotes the discontinuity across the surface interface, i.e. $[X]^{+}_{-}=[X]^{+}|_{\Sigma}-[X]^{-}|_{\Sigma}$. According to the Ref. \cite{Lobo:2005zu}, the momentum flux term $F_{\mu}$= $T_{\mu\nu}U^{\nu}$ in the right hand side corresponds to the net discontinuity across the shell.

Consequently, the energy-conservation equation on the shell $\nabla^{i}S^{i}_{i}$ can be written as $S^{i}_{\tau|i}$ =$\dot{\sigma}+2\dot{a}(\sigma+\mathcal{P})/a$ and the energy flux is given by
\begin{equation}
\left[T_{\mu\nu}e^{\mu}_{(j)}n^{\nu}\right]^{+}_{-}= - \frac{\left(\rho+p_r\right)\dot{a}\sqrt{1-2m/a+\dot{a^2}}}{1-2m/a}.
\end{equation}
One may deduced $\rho$ and $p_r$ from Eqs. (\ref{eq9}-\ref{eq10}), respectively, evaluated at the junction radius, $a$. Now, using the above relationship 
the conservation identity is much more appealing form
\begin{equation}
\sigma'= -\frac{2}{a}(\sigma+\mathcal{P})+ \Theta,
\end{equation}
where $\Theta$ is defined as 
\begin{equation}\label{th24}
\Theta = -\frac{1}{4\pi a} \frac{m'(1+ \omega)}{(a-2m)}\sqrt{1-2m/a+\dot{a^2}}.
\end{equation}
After some algebraic manipulations, from Eq. (\ref{eq7}), we then obtain
\begin{equation}\label{th25}
\Theta = -\frac{(w+1) \left(r^4+3 r^2 R^2\right)}{8 \pi  a \left(r^2+R^2\right) \left(a \left(r^2+R^2\right)-r^3\right)}\sqrt{1-2m/a+\dot{a^2}}.
\end{equation}
For $\omega = -1$, the Eq (\ref{th25}) reduces to zero, which relate to the solution obtained in Ref \cite{Visser:2003ge}.

We would also like to emphasize that for closing the system of equations we need an equation of state (EoS) $\mathcal{P}=\mathcal{P}(\sigma)$. Such an EoS would embrace the microphysics of the matter inside the shell. For instant, gravastar model has
an interior de-Sitter spacetime surrounded by a thin shell of ultra-stiff matter with an equation of state $p = \rho$, which is again matched to an exterior Schwarzschild vacuum solution.

\begin{table*}
\small
\begin{center}{TABLE 1: Numerical values of physical parameters for the different compact stars for $\omega$} \label{Table11-1}
\begin{tabular}{ccccccc}
 \hline
 Compact  & Observed Mass  & Predicted  & $\rho_{c}$  & $\rho_{s}$ & Surface & $\frac{2M}{R}$  \\
 
 Stars & $(M_{\bigodot})$ & Radius (Km) & $(gm/cm^{3})$ &  $(gm/cm^{3})$ & Redshift & \\
 \hline
PSR J1416-2230 (Demorest \textit{et al.} \cite{Demorest}) & 1.97 $\pm$ 0.04 & 11.083 $\pm$ 0.037 & 7.965 $\times 10^{14}$ & 5.695 $\times 10^{14}$ & 0.391 & 0.445\\
Vela X-1~ (Rawls \textit{et al.} \cite{R})  &   1.77 $\pm$ 0.08 & ~~10.852 $\pm$ 0.108  &  7.231 $\times 10^{14}$ & 5.371 $\times 10^{14}$ & 0.365 & 0.579\\
4U 1608-52~ (Gu$\dot{v}$er et al. \cite{Guver}) & 1.74 $\pm$ 0.14 & ~~10.811 $\pm$ 0.197 & 7.695 $\times 10^{14}$ & 5.975 $\times 10^{14}$& 0.342 & 0.436\\
Her X-1 (Abubekerov et al. \cite{Abubekerov})   & 0.85 $\pm$ 0.15 & ~~8.836 $\pm$ 0.481 & 8.014 $\times 10^{14}$ & 6.453 $\times 10^{14}$& 0.317 & 0.327\\
PSR J1903+327 (Freire \textit{et al.} \cite{F}) & 1.667 $\pm$ 0.021  & ~~10.703 $\pm$ 0.032 & 7.695 $\times 10^{14}$ & 5.832 $\times 10^{14}$& 0.215& 0.320 \\
 \hline
\end{tabular}
\end{center}
\end{table*}

\section{ Structural properties of compact objects}\label{sec4}

In order to examine more details about the stellar structure, we perform some analytical calculations and study the corresponding constraint for interior fluid sphere. Then, the effect of dark energy has been discussed to describe the mass-radius relation and explore massive stellar objects like massive pulsars, super-Chandrasekhar stars and magnetars, namely, PSR J1416-2230, Vela X-1 4U 1608-52, Her X-1 and PSR J1903+327 as given in Gangopadhyay \textit{et al} \citep{Gangopadhyay}. To be physically acceptable, the model should be free from any geometrical singularities i.e. energy density and pressure are regularity and finite at the center $r = 0$.

Owing to the several observations, the obtained solutions have been studied  and analyzed it's physical acceptability in terms of the star Vela~ X-1 (mass $1.77 \pm 0.08$ $M_{\bigodot}$ and radius $R = 10.852$ Km) by assuming it as the representative of compact star candidates. On substituting these values into Eqs. (\ref{eq9}-\ref{eq10}), we plot the dependence of energy density $\rho$ vs $r$ in Fig \textbf{2}. It may be pointed out here that energy density is finite inside the stellar interior and monotonic decreasing function of $r$. Since we have chosen dark energy EoS then it is obvious that radial pressure $p_r$ is negative inside the stellar interior. In Fig. \textbf{2} (right panel), we plot the anisotropic factor inside the stellar interior  for $\omega<-1/3$. We notic from the figure that $\Delta>0$  for $\omega = -0.35$ i.e. in the phantom region anisotropic force directed outward.

\subsection{Energy Condition}

We are interested here to investigate the energy conditions in terms of the components of
the energy-momentum tensor. To begin with, we consider different energy conditions, namely, null energy condition (NEC), weak energy condition (WEC), strong energy condition (SEC) and dominant energy condition (DEC) for the compact star candidate Vela~ X-1 (as the representative of other stellar model)- read as follows:
\begin{equation}\label{eq26}
(i) ~~~\text{\textbf{NEC}}: \rho(r)+p_r(r) \geq  0,\\
\end{equation}
\begin{equation}
(ii) ~~~\text{\textbf{WEC}}: \rho(r)+p_r(r) \geq  0~~ \text{and} ~~\rho(r) \geq  0,
\end{equation}
\begin{equation}
(iii) ~~\text{\textbf{SEC}}: \rho(r)+p_r(r) \geq  0~~ \text{and} ~~\rho+p_r(r)+p_t(r) \geq  0,
\end{equation}
\begin{equation}\label{eq29}
(iv) ~~~\text{\textbf{DEC}}: \rho(r)> \lvert p_r(r)\lvert ~~ \text{and} ~~\rho (r) > \lvert p_t(r)\rvert.
\end{equation}
Utilizing the inequalities, the nature of energy conditions for the astral structure Vela X-1 has been studied. The behavior of these conditions are shown graphically. In Fig. \textbf{3}, we plot the left hand side of the above inequalities as a function of $r$, which shows that all energy conditions of our model are satisfied for $\omega= -0.35$.

\begin{table*}
\small
\begin{center}{TABLE II: Physical parameters of the strange star candidate Vela X-1 due to the different values of state parameter $\omega$ with mass 1.77 $\pm$ 0.08 (Rawls \textit{et al.} \cite{R})} 

\begin{tabular}{ccccccc}
 \hline \hline
 Values of  & Predicted  & $\rho_{c}$  & $\rho_{s}$ & Surface & $\frac{2M}{R}$  \\
 
 $\omega$  & Radius (Km) & $(gm/cm^{3})$ &  $(gm/cm^{3})$ & Redshift & \\
 \hline
- 0.37 & 10.736  & 7.221 $\times 10^{14}$ & 5.351 $\times 10^{14}$ & 0.357 & 0.546\\
- 0.41  & 10.652   &  7.131 $\times 10^{14}$ & 5.142 $\times 10^{14}$ & 0.348 & 0.529\\
- 0.47 &  10.611 & 7.057 $\times 10^{14}$ & 4.957 $\times 10^{14}$& 0.337 & 0.486\\
 \hline
\end{tabular}
\end{center}
\end{table*}

\begin{figure}[h!]
\begin{center}
\includegraphics[width=8.6cm]{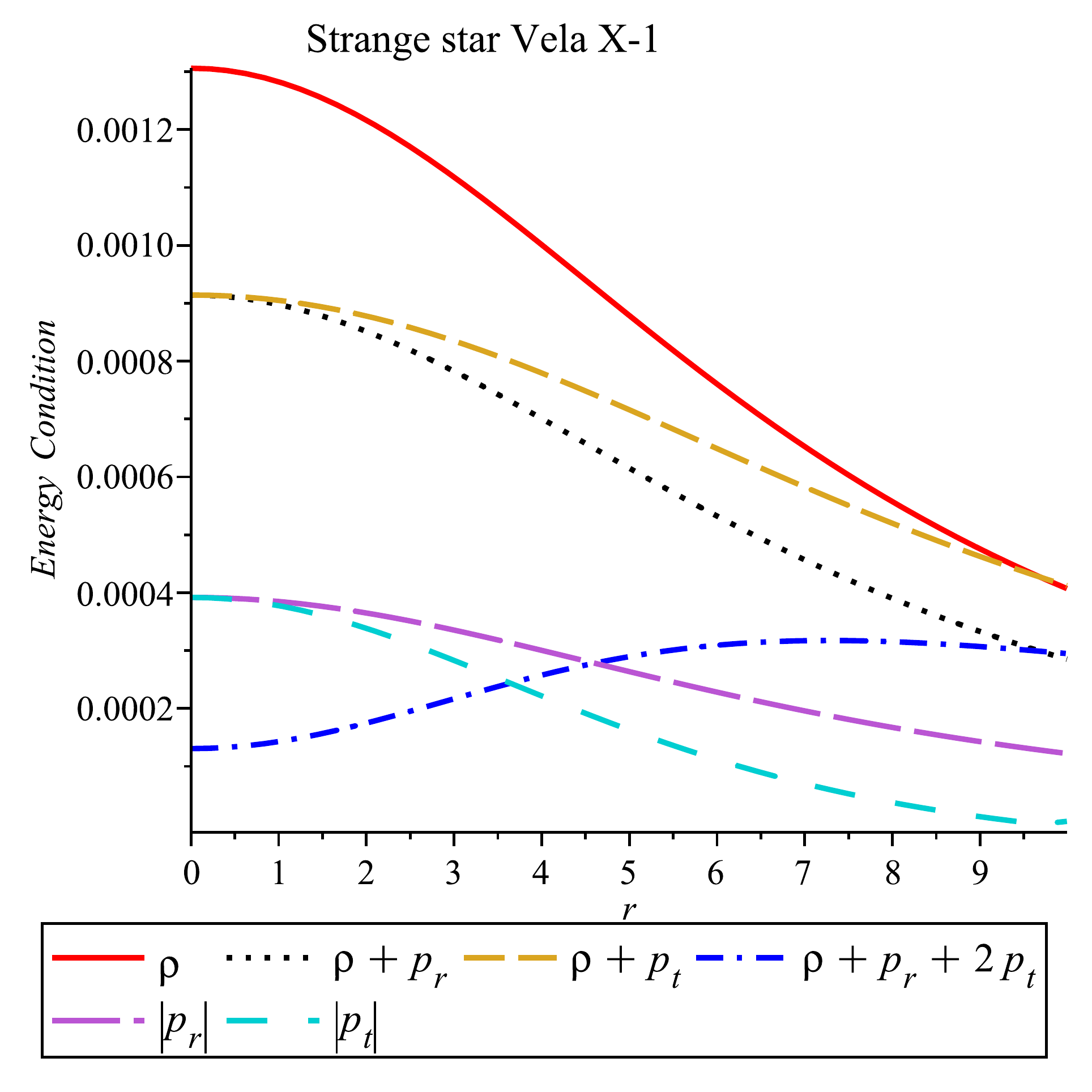}
\caption{The plot depicts the energy conditions for the inequalities Eqs. (\ref{eq26}-\ref{eq29}) as a function of the radial coordinate $r$. The curves for
null energy condition (NEC), weak energy condition (WEC), strong energy condition (SEC),  dominant energy condition (DEC) are shown in this figures, for the compact star Vela X-1.}
\end{center} 
\end{figure}

\begin{figure}[h!]
\begin{center}
\includegraphics[width=8cm]{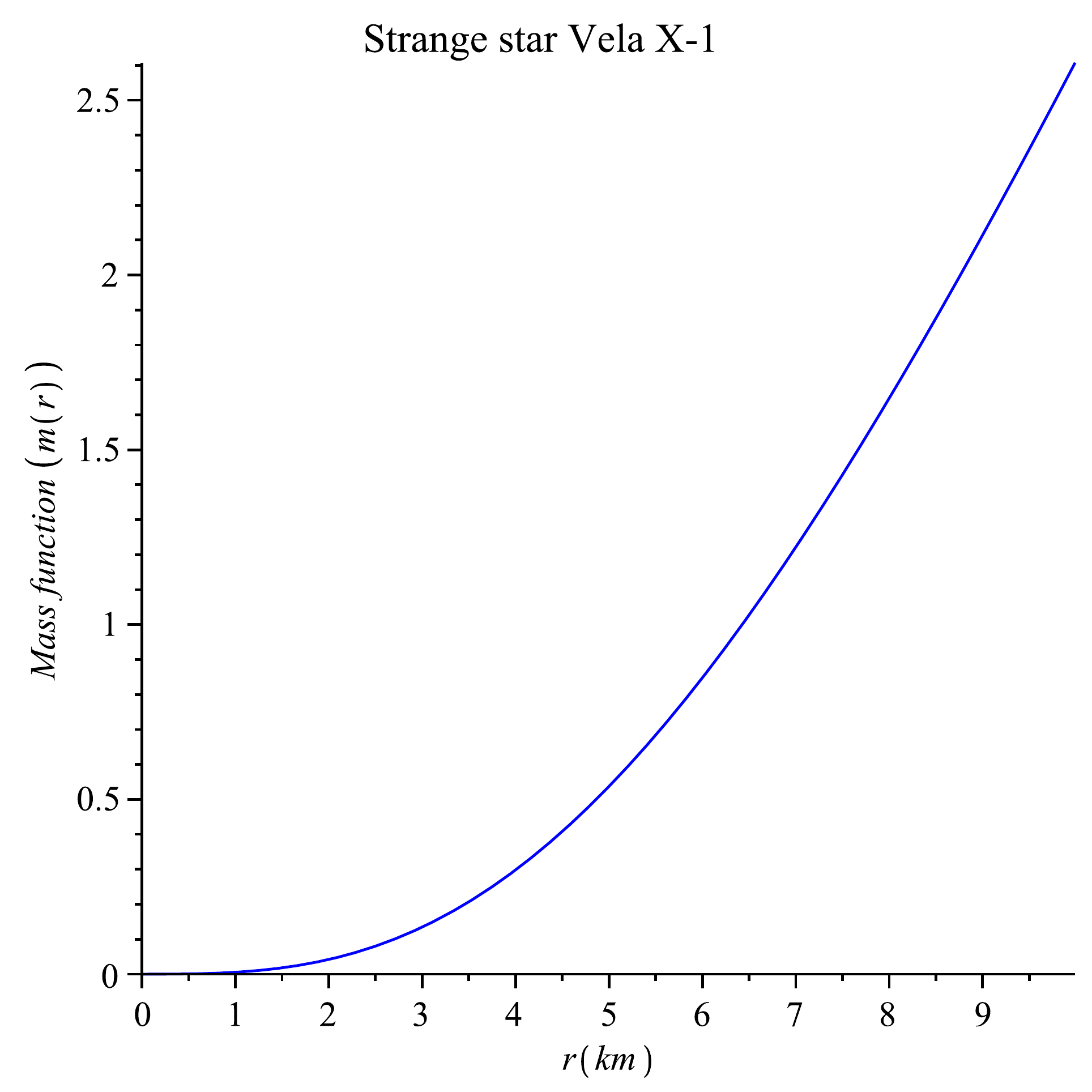}
\caption{The mass function $m(r)$ has been shown against $r$ of the Strange star candidate Vela X-1, which is
monotonic increasing within the radius of the star. }
\end{center} 
\end{figure}

\begin{figure}[h!]
\begin{center}
\includegraphics[width=8cm]{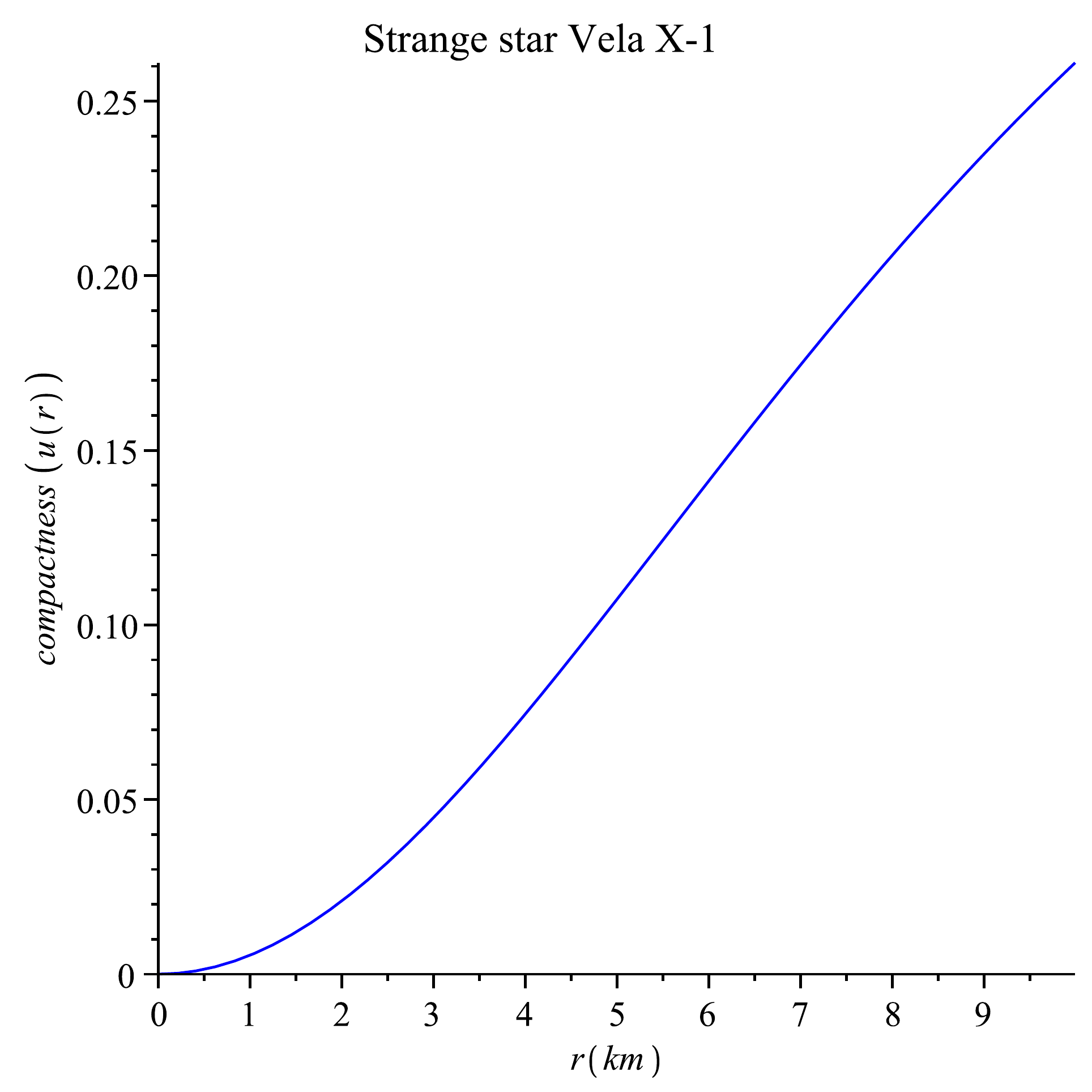}
\caption{The compactness of the strange star is shown against r, which
is monotonic increasing within the radius of the of strange star Vela~ X-1. }
\end{center} 
\end{figure}

\subsection{ Mass-radius relation and Surface gravitational red shift}
Here, we extend our analysis towards the effective mass-radius relation and surface gravitational red shift. For this purpose, the mass function within the radius of a compact star is given by
\begin{equation}\label{eq30}
m(r)=\int_0^{r}4\pi \rho r^{2}dr=\frac{r^{3}}{2(r^{2}+R^{2})}.
\end{equation}
We observe that the mass function $m(r)\rightarrow 0$ as $r\rightarrow 0$. Since the spherically symmetric perfect fluid configurations have the  allowable mass-radius ratio fall within the limit of $ 2M/R < 8/9$ (in the unit $c = G = 1$) \cite{Buchdahl}. Thus, putting the values considered for `Vela X-1' in Eq. (\ref{eq30}), we obtain the profile for mass function as shown in Fig. \textbf{4}. In order to check the condition we have plotted total mass normalized in solar mass, i.e. $M/M{_\odot}$  with the radius $R$ in Fig. \textbf{4}, for the specific value of $\omega$ = - 0.35. 
Indeed, it is shown that $m(r)$ is monotonic increasing function of radial co-ordinate and $m(r)>0$ for $0 <r <a$. To obtain the mass-radius relation for different compact stars, which are in general very closely equal to the observed values are tabulated in Table-\textbf{I}. Moreover, central density, surface density as well as surface redshift are tabulated Table-\textbf{I}
in describing the other compact configurations. From the observation one can see that the mass-radius relation for the different compact stars do not cross the proposed range of Buchdahl-Bondi inequality \cite{Buchdahl}.

\begin{figure}[h!]
\begin{center}
\includegraphics[width=8.6cm]{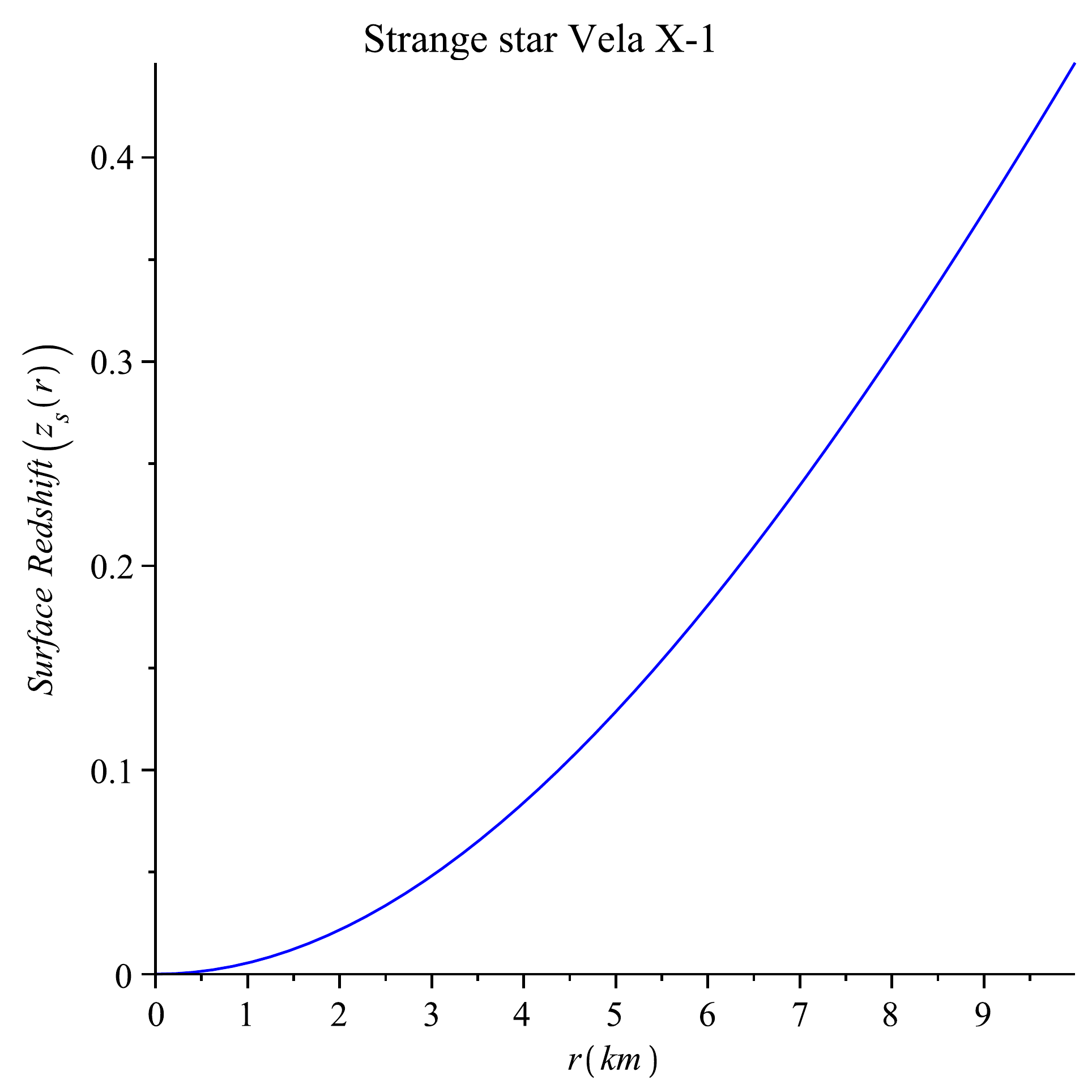}
\caption{Variation of the redshift function $z_s$
of the Strange star candidate Vela X-1 with respect to $r$ in km.}
\end{center} 
\end{figure}

Following the basic definition of the compactness of the star is obtained as
\begin{equation}\label{eq31}
u(r)=\frac{m(r)}{r}=\frac{r^{2}}{2(R^{2}+r^{2})}.
\end{equation}
We show the behaviour of the compactness $u(r)$ of the star in Fig. \textbf{5}. More precisely, the figure indicating that $u(r)$ is a monotonic increasing function of $r$,  and the redshift function $z_s $ of the compact star is given by
\begin{equation}\label{eq32}
1+z_s=(1-2u)^{-\frac{1}{2}}.
\end{equation}
Thus, the surface redshift function $z_s$ can be define as
\begin{equation}\label{eq33}
z_s=\sqrt{\left(\frac{R^{2}}{R^{2}+r^{2}}\right)}-1.
\end{equation}
In Fig. \textbf{6}, the surface redshift is shown for the compact star Vela X-1,  which is determined from the compactness parameter.
We also perform the maximum surface redshift ($z_s$) for different compact stellar configuration which is shown in Table \textbf{I} and Table \textbf{II}, respectively for different values of state parameter $\omega$. It is also of interest to observe that maximum value of the surface redshift, $z_s < 1$. In this sense, one may extract that obtained results  are compatible with the result obtained in \cite{hamity}.

\section{Results and Discussion}\label{sec5}
The driving forces behind scientific progress are contradictions between entrenched theories and new observations. Typically, dark energy models are based on scalar fields minimally coupled to gravity, is believed to responsible for the present accelerated expansion of the Universe. In fact, DE has opened up new possibilities in theoretical research for cosmology as well as astrophysical objects. In this article we present a formalism, based on dark energy EoS, describing a new class of static spherically symmetric stellar model satisfying all the physical realistic conditions.

The interior space-time metric is matched with the Schwarzschild exterior vacuum solution at the boundary. The constraints of the field equations are determined by the following conditions (i) the interior metric is describe by Finch-Skea type and (ii) the subsequent analysis is based on the dark energy EoS $p_r$ = $\omega \rho$. The motivation for implementing this model satisfies all the physical requirements like energy density, radial and transverse pressures which
are finite and regular at the centre. Therefore, it can potentially describe as a compact object which is neither a neutron stars nor a quark star. It is evident from Figs. \textbf{1} that the `gravitarional profile'  $g(r)>$ 0, indicating an inwardly gravitational attraction within the range of $-1/3<\omega<0 $, and $g(r)<0$ for $\omega<-1/3$, which indicate the gravitational repulsion. For a dark star solution it is necessary that the local acceleration due to gravity of the interior solution be repulsive in nature, so that the region where $g(r) >$ 0 is necessarily excluded. For value of $w = -0.35$, we consider Vela X-1 of mass 1.77 $M_{\bigodot}$ as the representative of compact star candidates. The results reported in Figs. \textbf{1}-\textbf{6}.  In particular, we pointed out that the central density is finite and maximum at the stellar interior (Fig. \textbf{2}), whereas the variation of anisotropy shown in Fig. \textbf{2} (right panel) is positive throughout the system for $-1<\omega <-1/3$. Hence, the direction of the anisotropic force is outward for our system.

Next, we matched our interior solution to the exterior Schwarzschild solution in presence of thin shell.  We set $G = c = 1$, while solving Einstein equations and plotting the graphs. Based on physical requirements and plugging the values of
$G$ and $c$ into the relevant equations, we calculate the numerical values of mass-radius relation for different compact stars. Physically the mass of the star is strongly dependent on its central density, and we know that high central density stars have lower gravitational masses. The obtained solutions are compared with the observed evidence for the existence of compact stars which is consistent with our model and the of mass-radius for different stars lies in the proposed range by Buchdahl \cite{Buchdahl} (The results reported in Table~\textbf{I} and ~\textbf{II}). To examine the nature of physical quantities, we consider a particular star, namely, Vela X-1 to investigate the physical properties. In order to investigate internal structure of the dark star in more details, we plot Figs. \textbf{3} - \textbf{6}, for energy conditions, mass-radius relations and surface redshift, respectively. In our model we have found the surface redshift ($z_s$) for the different compact stars are of finite values and vanishes outside of the star (see Table~\textbf{I} and \textbf{II} for more details), which typically fall within the proposed range in \cite{hamity}, are physically acceptable.

Finally, at very high mass measurement of about 2$M_{\bigodot}$ requires a really stiff
equation of state in neutron stars, and DE may be used as a possible candidate to study observable compact astrophysical objects.

\section*{Acknowledgement}
 MKJ acknowledge continuous support and encouragement from the administration of University of Nizwa. A. P also thanks the IUCAA, Pune for the support under visiting associateship programme where a part of this work has been done. We would also
like to thank the referee for his or her valuable comments.

\end{document}